\documentstyle[aps,preprint,tighten,floats,epsf,rotate]{revtex}

\begin{document}
\draft

%
\title{Strings with a confining core in a Quark-Gluon Plasma}
\author{Biswanath Layek \footnote{e-mail: layek@iopb.res.in},
Ananta P. Mishra \footnote{e-mail:apmishra@iopb.res.in}, and 
 Ajit M. Srivastava \footnote{e-mail: ajit@iopb.res.in}}
\address{Institute of Physics, Sachivalaya Marg, Bhubaneswar 751005, 
India}
\maketitle
\widetext
\parshape=1 0.75in 5.5in
\begin{abstract}
 
 We consider the intersection of N different interfaces interpolating
between different $Z_N$ vacua of an SU(N) gauge theory using the
Polyakov loop order parameter. Topological arguments show that at
such a string-like junction, the order parameter should vanish,
implying that the core of this string (i.e. the junction region of all 
the interfaces) is in the confining phase. Using the effective potential 
for the Polyakov loop proposed by Pisarski for QCD, we use numerical 
minimization technique and estimate the energy per unit length of the 
core of this string to be about  2.7 GeV/fm at a temperature about twice the
critical temperature. For the parameters used, 
the interface tension is obtained to be about 7 GeV/fm$^2$. Lattice
simulation of pure gauge theories should be able to investigate properties 
of these strings. For QCD with quarks, it has been discussed in the 
literature that this $Z_N$ symmetry may still be meaningful, with quark 
contributions leading to explicit breaking of this $Z_N$ symmetry. With 
this interpretation,  such {\it QGP} strings may play important role
in the evolution of the quark-gluon plasma phase and in the dynamics of
quark-hadron transition.

\end{abstract}
\vskip 0.125 in
\parshape=1 0.75in 5.5in
\narrowtext

\section{Introduction}

 Physics of quark-gluon plasma (QGP) and the quark-hadron phase transition
has become a very active area of research in recent years. There are 
several reasons for this. The most important motivation comes from the
ongoing, and upcoming, relativistic heavy-ion collision experiments
where it is widely believed that a hot dense region of QGP will be
created. This will provide the opportunity for a controlled study
of high temperature phase of a theory as rich as QCD, as well as
the dynamics of phase transition in a relativistic quantum field
theory. Such studies are especially important for the early universe
where the universe was in the QGP phase until it was about few 
microseconds old. Many studies have been carried out about possible
observational implications of the quark-hadron phase transition
in the early universe. Further, lattice results have given a control
on the understanding from the theory side and the stage is set for
confronting observations with lattice results for deeply
non-perturbative regime of the quark-hadron phase transition.

  This richness of QCD is manifested in non-trivial phases and 
structures allowed by the theory in various regimes of temperature,
chemical potential, and other control parameters (some of which
may be primarily of theoretical importance, such as quark masses).
Even the perturbative regime of high temperature QGP phase has
non-trivial vacuum structure as seen by using the expectation 
value of the Polyakov loop $l(x)$ as the order parameter for the
confinement-deconfinement phase transition \cite{plkv}. This order 
parameter transforms non-trivially under the center $Z(3)$ of the color 
SU(3) group and is non-zero above the critical temperature $T_c$. This 
breaks the global $Z(3)$ symmetry spontaneously above $T_c$, while
the symmetry is restored below $T_c$ in the confining phase where
this order parameter vanishes. 

  In the QGP phase, due to spontaneous breaking of the discrete $Z(3)$
symmetry, one gets domain walls (interfaces) which interpolate between
different $Z(3)$ vacua. The properties and physical consequences of 
these $Z(3)$ interfaces have been discussed in the literature \cite{zn}. 
(Though, we mention that it has also been suggested  that these 
interfaces should not be taken as physical objects in the Minkowski 
space \cite{smlg}. Similarly, it has also been subject of discussion
whether it makes sense to talk about this $Z(3)$ symmetry in the presence
of quarks \cite{qurk1}.) The presence of quarks can be interpreted as
leading to explicit breaking of $Z(3)$ symmetry, lifting the degeneracy 
of different $Z(3)$ vacua \cite{qurk2,psrsk,psrsk2}. In this approach, 
with quarks, $Z(3)$ interfaces become unstable 
and move away from the region with the 
unique true vacuum. We will take this interpretation to be valid for the 
case with quarks.  Though our main discussion will be for the pure gauge
theory which we discuss first. Later we will comment on the situation 
with quarks.

 We will consider specific configurations of $Z(3)$ interfaces, namely
the intersection of all three different interfaces in the high temperature 
deconfining phase. (For general SU(N)
case one will consider the intersection of N different $Z(N)$ interfaces.)
Topological arguments then show that at the line-like intersection of these
interfaces, the order parameter $l(x)$
should vanish. This leads to a topological string configuration with 
core of the string being in the confining phase. It is important to
note that such a string has exactly reverse physical behavior compared to
the standard QCD string. The QCD string exists in the confining phase,
connecting quarks and antiquarks, or forming baryons, glueballs etc.
Inside the QCD string energy density is high enough, and distance scales
small enough, that the core region behaves as  a deconfined region.
In contrast the string we are discussing exists in the high temperature
deconfined phase. Its core is characterized by restored $Z(3)$
symmetry, implying that it is in the confined phase. To differentiate
it with the standard QCD string, we call this string configuration
the QGP string. It is also important to note that although the standard
QCD string breaks by creating quark-antiquark pairs, the QGP string
cannot break as it originates from topological arguments. This QGP 
string, thus should either form closed loops, or it should end at the
boundary separating the deconfined phase from the confined phase.

 To estimate the physical properties of such a string configuration for QCD,
we use the effective potential proposed by Pisarski \cite{psrsk,psrsk2}
(see, also ref. \cite{veff}) for the Polyakov loop $l(x)$. Using this 
Polyakov loop model, we estimate the energy per unit length of the QGP 
string to be about 2.7 GeV/fm at a temperature about $2T_c$. 
The interface tension is found to be about 7 GeV/fm$^2$ at this
temperature.

  It is clear that the structure of this QGP string is similar to the
standard axionic string which forms at the junction of axionic domain walls 
\cite{axion}. The consequences of the QGP string therefore can be
determined following those discussions. Presence of quarks can
then be taken in terms of explicit $Z(3)$ symmetry breaking. This will
lead to decay of $Z(3)$ interfaces along with rapid decay of the 
associated strings.  One important difference with the axionic strings is
that the axionic strings are supposed to be produced at the Pecci-Quinn
symmetry breaking transition, with symmetry broken in the low temperature 
phase. Thus the standard picture of string formation in a phase transition
is applicable. In contrast, for the QGP strings, symmetry is always broken
in the high temperature phase, it gets restored below $T_c$, the
QCD transition temperature. The formation of these QGP strings thus
will be of a different nature. Further, this 
string network will melt away below $T_c$. Another difference between 
the present case and the standard axionic models is that, as we will see
below, for the axionic models the central bump in the effective potential 
is higher than the barrier  between different $Z(N)$ vacua. In those 
axionic models, one can therefore restrict the order parameter $|\phi| 
\sim \phi_0$, where $\phi_0$ is approximately the value of the order 
parameter corresponding to the absolute minimum of the effective potential.
This  reduces the problem effectively to that for a scalar field with  a 
disconnected vacuum manifold. The solution for kink solitons are well known 
in such models. In contrast, for the present case  the situation
is reversed. Here the barrier between different $Z(N)$ vacua is much
higher than the central bump. Thus, here the problem cannot be reduced to 
that of the standard scalar kink soliton and as we will explain, the 
discussion of appropriate domain wall solution becomes much more involved.

 Presence of such strings should have strong effects on the properties of
the QGP phase as well as on the dynamics of the quark-hadron phase
transition. With pre-existing string network with confining core,
the transition to the confining phase should begin from regions near
the string. This is certainly true for a first order transition, where
the string network will make heterogeneous nucleation, with bubbles
of confining phase forming at these QGP strings, more dominant compared
to the conventional homogeneous nucleation. Even for a second
order transition, the transition may not remain uniform, and may
proceed from the strings outward. We hope to discuss the effects of
such a modified picture of phase transition on baryon inhomogeneity
generation for the early universe as well as for relativistic heavy
ion collision experiments in a future work.

 The paper is organized in the following manner. In section II, we
briefly review the Polyakov loop model of Pisarski and discuss the
topological arguments leading to the existence of the QGP string.
In section III, we discuss the numerical technique and the properties 
of the $Z(3)$ interface. Section IV discusses junctions
of these interfaces, i.e. the profile of the QGP string. We 
also comment here on the effect of quarks on these solutions.
Section V presents conclusions.
 
We mention here that the string-like junction of these $Z(N)$ interfaces 
for an SU(N) theory has been discussed in ref.\cite{mnpl}, but with a
completely different interpretation \cite{mnpl}. In ref.\cite{mnpl} these
intersections are identified with the space-like world lines of 
color magnetic monopoles in the context of
confinement by monopole condensation. 

\section{The Polyakov loop model}

 As we mentioned above, we will focus on pure SU(N) gauge theory and
later discuss the case with quarks. In this case, an order parameter for 
the confinement-deconfinement phase transition is the Polyakov loop
$l(x)$ which is defined as,

\begin{equation}
l(x) = \frac{1}{N} tr \Bigl(P exp\Bigl( ig \int^\beta_0 A_0(x,\tau) 
d\tau \Bigr) \Bigr) .
\end{equation}

 Here $P$ denotes path ordering, $g$ is the gauge coupling, $\beta
= 1/T$, with $T$ being the temperature, $A_0(x,\tau)$ is the time
component of the vector potential at spatial position $x$ and Euclidean
time $\tau$. $l(x)$ is thus a complex scalar field. Under a global 
$Z(N)$ symmetry transformation, $l(x)$ transforms as,

\begin{equation}
l(x) \rightarrow exp\bigl(\frac{2\pi i n}{N}\bigr) l(x), 
~~ n = 0,1,..(N-1) .
\end{equation}

 For temperatures above the critical temperature $T_c$, in the deconfining
phase, the expectation value of the Polyakov loop $l_0 = <l(x)>$ is non-zero
corresponding to the finite free energy of isolated test quarks. This
breaks the $Z(N)$ symmetry spontaneously. At temperatures below $T_c$,
in the confining phase, $l_0$ vanishes, thereby restoring the $Z(N)$
symmetry \cite{plkv}. We now restrict to the case of QCD with $N = 3$ 
and take the effective theory for the Polyakov loop as proposed by Pisarski
(see ref. \cite{psrsk,psrsk2} for details), given by the following 
effective Lagrangian density,

\begin{equation}
L = \frac{N}{g^2} |\partial_\mu l|^2 T^2 - V(l) .
\end{equation}

 Here, $N = 3$ and $V(l)$ is the effective potential for the Polyakov
loop given by,

\begin{equation}
V(l) = \Bigl( -\frac{b_2}{2}|l|^2 - \frac{b_3}{6}\bigl(l^3 + (l^*)^3 
\bigr) + \frac{1}{4} \bigl(|l|^2 \bigr)^2 \Bigr) b_4 T^4 .
\end{equation}

$l_0$ is then given by the absolute minimum of $V(l)$.  Normalization
of $l(x)$ is chosen such that $l_0 \rightarrow 1$ as $T \rightarrow
\infty$.  Values of various parameters in Eqs.(3),(4) are fixed in 
ref.\cite{psrsk2} by making correspondence to lattice results. We make 
the same choices, and give those values below.

The values of various parameters in Eq.(4) are fixed to reproduce
lattice results \cite{lattice} for pressure and energy density
of pure SU(3) gauge theory. For pure gauge theory we will use the same
parametrization as is chosen in ref.\cite{dumitru1} where the coefficient
$b_3$ and $b_4$ has been taken as,  $b_3=2$ and $b_4 = 0.6061$. We will take
the same value of $b_2$ for real QCD (with three massless quark flavors),
while the value of $b_4$ will be rescaled by a factor of $47.5/16$ to 
account for the extra degrees of freedom relative to the degrees of 
freedom of pure gauge theory, as in ref.\cite{dumitru1}.  The temperature 
dependent coefficient $b_2(T)$ has been taken 
from ref.\cite{dumitru1,psrsk2} 
which is expressed in terms of the ratio $r$ ($=T/T_c$) as, 
$b_2(r) = (1-1.11/r)(1+0.265/r)^{2}(1+0.300/r)^{3}-
0.487$. With the coefficients chosen as above, the expectation value
of the order parameter approaches to $x =b_3/2 + 
\frac{1}{2}\sqrt {b_3^2 +4~b_2(T=\infty)}$ for temperature $T \rightarrow 
\infty.$ As in ref.\cite{psrsk2}, we will use the normalization such that 
the expectation value of the order parameter $l_0$ goes to unity for
temperature $T \rightarrow \infty$, hence the fields and the coefficients 
are rescaled as $l \rightarrow l/x, b_2(T) \rightarrow b_2(T)/x^2,b_3
\rightarrow b_3/x$ and $b_4 \rightarrow b_4~x^4$ to ensure proper
normalization of $l_0$.

 By writing $l = |l| e^{i\theta}$ we see that the $b_3$ term in Eq.(4)
gives a $cos(3\theta)$ term, leading to $Z(3)$ degenerate vacua for 
non-zero values of $l$, that is for $T > T_c$. With the choice of
parameters as above, the value of $T_c$ is $\sim 182$ MeV. The $Z(3)$
interface solution will correspond to a planar solution (say in
the x-y plane) where $l$ starts at one of the minimum of $V(l)$ at
$z = - \infty$ and ends up at the other minimum of $V(l)$ at $z = + \infty$.
These $Z(3)$ interfaces have been extensively discussed in the 
literature \cite{zn}.

  Consider now the junction of all three different $Z(3)$ interfaces, 
say with the line like junction being along the z axis. The interfaces then 
will be perpendicular to the x-y plane. It is immediately
clear that as one considers a closed loop in the physical space encircling  
the z axis, then one encircles $l = 0$ point in the complex $l$ plane.
It is then obvious from the continuity of $l$ that $l$ must vanish along
the z axis. This is the standard argument for topological string solutions
(more specifically axionic strings, though situation here is somewhat 
different due to large barrier height, as discussed below). These are 
topological strings and hence cannot break. Note that, in contrast the 
standard QCD string can break by creating quark-antiquark pairs.

The potential 
given in Eq.(4), leads to a weak first  order transition. We show the plot 
of  $V(l)$ in $\theta = 0$ direction in Fig.1a for a value of
temperature $T = 185$ MeV. This shows the metastable
vacuum at $l = 0$. To see the $Z(3)$ structure of the vacuum we plot
the potential $V(l)$ as a function of $\theta$ for fixed $|l| = l_0$,
where $l_0$ corresponds to the absolute minimum of $V(l)$. This is
shown in Fig.1b. An important thing to note from Fig.1a,b, is that the
height of the barrier between different $Z(3)$ vacua is much higher
than the height of the barrier between the metastable vacuum at $l = 0$
and the true vacuum. This situation is in complete contrast to the standard
axionic models where the central bump is higher than the barrier 
between different $Z(N)$ vacua. In those axionic models, one can therefore
restrict $|l| \sim \l_0$ which reduces the problem effectively to that 
with a scalar field with  a disconnected vacuum manifold. The solution for 
kink solitons are well known in such models.

  Such a solution cannot be found here, as the height of barrier between
the $Z(3)$ vacua is much higher than the central bump. At high temperatures
the ratio of the heights of the $Z(3)$ barrier and the central bump reduces,
but it still remains much larger than 1. In Fig.2a we show
the surface plot of the potential $V(l)$ in Eq.(4) in the complex $l$ plane 
for $T = 400 MeV$. Plot range in the $V$ axis is restricted to make 
the $Z(3)$ barriers and the central bump distinctly visible. To
compare this situation with a standard axionic model case (with same
$Z(3)$ symmetry), we show in Fig. 2b the potential in Eq.(4) with the
coefficient $b_2$ multiplied by 100. This raises the central bump
higher than the $Z(3)$ barriers. For a potential like in Fig.2b the
$Z(3)$ domain wall solution is physically clear as the order parameter 
interpolates between the different $Z(3)$ vacua through the (smaller)
barrier in the valley. In contrast, in Fig.2a, the lowest potential energy 
path from one $Z(3)$ vacuum to another goes through the origin $l = 0$.
It may give the impression that the $Z(3)$ interface will have $l = 0$
in the middle of the interface. If that was true then it will imply that
these $Z(3)$ interfaces have confining regions in the middle, which will
not be in agreement with other studies of $Z(3)$ interfaces for QCD 
\cite{zn}. However, as we will discuss in the next section, the actual 
domain wall, which is a solution of the field equations, cannot have $l = 0$
inside it. Still, a potential like in Fig.2a, makes it impossible to reduce
the Lagrangian of Eq.(3) to an effective Lagrangian with kink solutions
(as in the case of axionic models), and one has
to find numerical techniques to determine the appropriate interface
solution. We discuss this in the next section.

\section{Numerical techniques and the $Z(3)$ interface profile}

  First let us see general properties of the interface solution for the
effective potential shown in Fig.2a. We are interested in time
independent solutions. (This is in the absence of quarks. With quarks,
$Z(3)$ symmetry will be explicitly broken, so there will not be any
time independent interface or string solutions.) For a planar wall 
in the x-y plane we can write down the field equation as

\begin{equation}
\frac{d^2 l_i}{dz^2} = \frac{g^2}{NT^2} 
\frac{\partial V(l_1,l_2)}{\partial l_i} ,
\end{equation}

\noindent where we have written $l = l_1 + i l_2$. Using generalization of
standard techniques \cite{clmn}, we interpret $z$ as time, and $l_1$ and 
$l_2$ as the two dimensional position space for a particle which is moving
under the influence of potential = $ - V(l_1,l_2) (\frac{g^2}{NT^2})$. 
Domain wall solution will then correspond to the particle trajectory 
which for $z \rightarrow - \infty$ approaches one of the minima of $V(l)$,
while for $z \rightarrow + \infty$ it approaches a different
minima of $V(l)$. 

  Fig.3 shows the plot of the inverted potential, i.e. $ - V(l_1,l_2)$.
Minima of $V$ now become maxima of $-V$. Again, to show the shape
of the potential clearly, we have restricted the range of plot for
negative values. As we mentioned above, the domain wall solution will 
correspond to the particle trajectory starting at the top of one of the 
hills, say, at $P$, and ending at the top of another hill, say at $Q$. 
With this picture it becomes immediately obvious that the domain wall
solution cannot go through $l = 0$. The particle starting at $P$, and
rolling down to $l = 0$ cannot turn back to end up at $Q$. It will 
rather go to the other side and roll away downward, as shown by the 
dashed curve in Fig.3. To end up at $Q$, the trajectory must loop back
before reaching $l = 0$, as shown by the solid curve in Fig.3. 
Thus, $l$ must remain non-zero inside the domain wall. 

  What we have described above is a simple generalization of the basic
idea of how domain wall solution is found for a real scalar field. For 
real scalar field, one can numerically determine the solution of the field 
equation by tuning up the initial starting point of the particle 
(i.e. the field value), depending on whether the solution overshoots,
or undershoots at large time (i.e. large $z$) \cite{clmn}. However, for 
complex $l$, this tuning up of the initial condition becomes very difficult
to determine depending on the nature of large time (large $z$) 
behavior of the solution. One will need to tune up both components
$l_1$ and $l_2$ appropriately, depending on the details of large
$z$ behavior, and it becomes very difficult to develop suitable
criterion for this. This is certainly an interesting problem to
develop suitable numerical scheme for numerically solving the 
differential equations to determine the domain wall
solution for this type of potential.

  In the absence of such a technique to determine the solution of the
differential equation, we resort to numerical minimization of
the energy to determine the appropriate solutions. Such a method is
in general difficult to implement as one needs to be reasonably
certain that one has not found a local minimum of the energy
functional. The only way is to try different initial configurations
with varying lattice sizes, and variational parameters, and see whether
one gets the same final configuration. We have carried out such tests
in our numerical minimization and we believe our results are
trustworthy from this point of view.

  For time independent case, the energy density from Eq.(3),(4) is,

\begin{equation}
{\cal E} =  \frac{N T^2}{g^2} |\bigtriangledown l|^2 +  V(l) .
\end{equation}

  For the planar domain wall solution (say in x-y plane), this energy 
density is integrated along the z axis to get the energy per unit
area $E_W$ of the wall. For string solution along z axis (discussed in the 
next section), this energy density is integrated in the x-y plane to
get the energy per unit length $E_S$ of the string. Numerical minimization
of $E_W$ (and $E_S$ respectively) is carried out as described below. 

 For the energy minimization, we have used a code as was used in 
ref.\onlinecite{emin}. To determine the domain wall solution we only
need to consider the profile of $l$ in one dimension (along $z$), so we 
restrict the two dimensional simulation of ref.\cite{emin} to one 
dimension. We fix the values of $l$ at the two boundaries of the one 
dimensional lattice as $l = l_{01}$ and $l = l_{02}$ where $l_{01}$ 
and $l_{02}$ are the values of $l$ corresponding to the two distinct 
minima of $V(l)$. For the  intermediate lattice points we use an 
interpolating configuration between these two values. For small
physical size of the lattice we use linear interpolation from one
boundary to the other boundary, while for a large physical size of
the lattice (compared to wall thickness) we use linear interpolation 
in a smaller, central portion of the lattice. In both situations we 
get the same final configuration for the interface. (If we use linear 
interpolation for the entire large lattice also then the minimization 
program gets trapped in some local minimum of energy converging to a 
configuration which has much larger energy.) We use a large lattice with 
10$^4$ points, with lattice spacing being 0.002 fm. The physical size 
of the one-dimensional lattice is then 20 fm. In this case we used
linear interpolation for the middle 10 fm portion of the lattice for
the initial configuration. Field configuration is 
then fluctuated at each lattice point, while fixing the
boundary points, and energy is minimized. The configuration with the 
lowest value of energy is finally accepted (when the energy almost settles
down to a definite value) as the correct profile of the 
interface and corresponding energy is taken as the energy (per unit area) 
of the domain wall. In the following we explain the essential aspects 
of this energy minimization technique \cite{emin}.

 We have used over relaxation technique for energy minimization, as in
ref.\cite{emin}, which we have found to be very efficient for our case. 
This consists in first determining
the most favorable fluctuation in the field at a given site by fluctuating
field there and considering the change in the energy density. The most
suitable fluctuation corresponds to the minimum of the parabola which
passes through these values of energy densities (corresponding to
fluctuated values of the field). Then the
actual change in the field is taken to be larger (by a certain factor)
than this most suitable fluctuation. We have found that changing this
factor in the range of 0.01 - 0.03 worked best for our case.

  The minimization code has been tested with two dimensional lattice
for finding the configuration of a standard U(1) global string, with a 
complex scalar order parameter $\phi$ \cite{emin}. Initial configuration 
of $\phi$ is chosen to be a winding number one, azimuthally symmetric 
configuration with some initial function for  $|\phi(r)|$. We then 
minimize the energy and determine $\phi(r)$ which gives the lowest
energy configuration. It is found that even if the initial profiles
for $\phi(r)$ prescribed are very different (for example we have
tried out triangular form for $\phi(r)$) , after about 200
iterations, $\phi(r)$ converges to the exact solution as obtained
by numerically solving the field equations using a Runge- Kutta 
algorithm of fourth order accuracy.

  Fig.4 shows the profile for the domain wall solution for $T = 400 MeV$.
The surface tension of the wall in this case is found to be about 
7 GeV/fm$^2$. Note that $l$ remains non-zero in the profile of the wall, as
we argued above. As $T$ approaches $T_c$, the heights of the barriers 
between different $Z(3)$
vacua become much higher compared to the central bump (in Fig.2a). In
such a situation, as one can see from Fig.3, in order to end up at Q,
starting from P, the particle trajectory will have to loop back from
a point which gets closer to the origin. This is indeed what we see
in our numerical simulation also. We find that as $T$ approaches $T_c$,
the value of $l$ in the middle of the domain wall profile becomes
very small, but it always remains non-zero. We use $T = 400$ MeV to get 
the profile of the domain wall (and the QGP string) to show distinctly 
non-zero $l$ profile for the wall. We mention here that, as mentioned in 
ref. \cite{psrsk2}, the parameter values used here are not valid for high 
temperatures. Our purpose here is not to be very precise about the numbers 
we get but about the qualitative aspects of the solutions we get. Also,  
the energies etc. we get may be correct up to  factors of order unity.

\section{Junctions of $Z(3)$ interfaces, the string profile}

  We now consider a configuration which corresponds to the junction of
three different $Z(3)$ interfaces. As we discussed above, by considering
a closed loop in the physical space encircling this junction, 
we see that the order parameter encircles $l = 0$ point in the 
complex $l$ plane. From the continuity of $l$, it then follows 
that $l$ must vanish along the line forming the junction of the
interfaces. This, therefore, leads to a topological string configuration 
whose core is in the confining phase. 

To determine the profile of this QGP string, we use the numerical 
minimization  as described in the previous section, with a two
dimensional lattice. We have used a 600 $\times$ 600 lattice,
with lattice spacing being 0.01 fm. The physical size of the lattice
is then 6 fm $\times$ 6 fm. The string is taken to be perpendicular to 
the lattice, with the lattice giving a cross-section of the string, 
as well as the interfaces attached to it. We start with a trial 
configuration which has isotropic variation of $\theta$ (as
appropriate for the conventional U(1) string), with the magnitude
of $l$ vanishing at the center of the lattice. For the trial configuration
we take the radial profile of $|l|$ such that it is zero at the center
of the lattice and increases to the vacuum value of $l$ exponentially with 
a typical distance scale of few fm. We know that for the $Z(3)$ string,
$\theta$ variation will not remain isotropic,  it will become
concentrated in the three domain walls whose junction will be the
string. Thus we cannot fix the field at the boundary of the two dimensional 
lattice during minimization procedure. If we do not fix field anywhere 
and carry out the minimization,
then string leaves the lattice due to asymmetric interface lengths (for
a square lattice which we use). To handle this problem, we fix the
center of the string \cite{emin}. The center of
the string (where $l = 0$) is chosen to lie at the middle of an elementary 
lattice square. We then fix $l$ at the four lattice points forming this
particular lattice square. Everywhere else $l$ is fluctuated, and
the energy is minimized to get lowest energy string profile. Since $l$ is
fixed only for a very tiny elementary square (with lattice step 
being 0.01 fm), it causes negligible error in the determination of
the correct string profile and string energy.

 Fig.5 shows the surface plot of $-l$ showing clearly the string, connected 
to three interfaces.  We thus see that despite very different barrier ratios
for the potential in Eq.(4) and the standard axionic case (as shown in
Fig.2a,2b), the string configuration is very similar to the standard
axionic string connected to the $Z(3)$ interfaces. Determination of
the energy of the string in this case becomes ambiguous due to the
contribution of the energy of the interfaces. To separate out the
string core energy contribution we use the following method. For the
two dimensional cross-section of string profile, we obtain
the net energy $E(r)$ within radius $r$ starting from the center of 
the string by integrating ${\cal E}$ in Eq.(6) for the configuration
of Fig.5 inside a circle of radius $r$, with the center of the circle
being at the center of the string.
$E(r)$ will get contribution from the core of the string
initially, but for large $r$ the contribution of interface energy will
dominate. We can then parametrize $E(r)$ as follows,

\begin{eqnarray}
E(r) = E_0(r) , ~~~  0 < r < r_0 \\
E(r) = E_0(r_0) + 3 \sigma (r - r_0), ~~~ r > r_0
\end{eqnarray}

Here, $E_0(r)$ denotes the core energy contribution which should be the
dominant contribution up to some distance $r_0$. Beyond $r_0$, the
linear contribution of interfaces becomes significant. By plotting
$E(r)$ vs. $r$, we can get $\sigma$ as well as the core energy $E_0(r_0)$.
Fig.6 shows this plot. We have fitted the large $r$ part of the plot
with a straight line. Its slope is found to be about 23 GeV/fm$^2$, giving 
the value of $\sigma \simeq 7.7$ GeV/fm$^2$ in reasonably good agreement 
with our numerical estimate for the wall tension given in the previous 
section. The core energy $E_0(r_0)$ is found to be about 2.7 GeV/fm.

 Let us now come back to the issue of quarks and the $Z(3)$ symmetry.
The effect of quarks on this $Z(3)$ symmetry and $Z(3)$ interfaces etc.
has been discussed in detail in the literature \cite{qurk1,qurk2}. It has 
been suggested that in the presence of quarks, the $Z(3)$ symmetry becomes 
meaningless, and there is no sense in talking about $Z(3)$ interfaces etc.
\cite{qurk1}. It has also been advocated in many papers, that one can take 
the effect of  quarks in terms of explicit breaking of $Z(3)$ symmetry
\cite{qurk2,psrsk,psrsk2}. In such a case, the interfaces will survive, 
though they do not remain solutions of time independent equations of motion. 
It has been argued in ref.\cite{psrsk2} that the effects of quarks in terms 
of explicit symmetry breaking may be small, and the pure glue Polyakov model
may be a good approximation. We will therefore assume
that the effects of quarks is either negligible, or it just
contributes explicit symmetry breaking terms which can make the
interface and the string solution time dependent, but not invalid.
 
   With the explicit symmetry breaking, the interfaces and string
will develop dynamics, for example, the interfaces will start moving
away from the direction where true vacuum exists. The string will also not
have three interfaces forming symmetrically around it, and hence will 
start moving in some direction. Such motions may cause important 
differences on long time behavior (as for the axionic strings), but
for short time it may be immaterial. This is because strings and
domain walls anyway move around after formation due to the fact
that they have large tensions and at the time of formation they 
almost never form in symmetric configurations. Thus the initial time
dynamics of these $Z(3)$ interfaces and QGP string may not be much 
distinct from the case when there is no explicit $Z(3)$ symmetry 
breaking. This should be the situation appropriate for relativistic
heavy-ion collisions. Of course the issue of initial time here is 
subtle as the strings exist in the high temperature phase, compared to 
the standard axionic strings which exist in the low temperature phase.
Effects of these strings
and their properties for pure gauge theories could be investigated
by lattice simulations. As elaborated above,
these strings, which are embedded in the QGP phase, have confining 
core.  Because of this these strings can affect the dynamics of
quark-hadron phase transition in important ways. It is possible that, 
as the transition temperature is approached from above, string cores 
may swell, and trigger the transition process. For a first order
transition, the string with its confining core will act as an ideal
site for bubble nucleation. This type of situation is often seen in
condensed matter systems. For example in a nematic liquid crystal
system when the system is heated back to the isotropic phase,
with strings existing in the nematic phase, bubbles of isotropic
phase nucleate primarily on top of the string defects. Same thing should
happen here also. As the confining core exists inside the QGP strings,
nucleation of a bubble on top of strings requires smaller free  energy
barrier to be overcome by thermal fluctuations. We thus expect that 
instead of homogeneous nucleation of bubbles one should get bubble 
nucleation happening on top of the entire QGP string network. We hope 
to investigate some of these possibilities in a future work. 

\section{conclusions}

 We have discussed special configurations of junctions of $Z(N)$ interfaces
for an SU(N) gauge theory and have shown that these correspond to 
topological strings which have confining phase in the core. Using the 
Polyakov loop model of Pisarski \cite{psrsk} for QCD, we have estimated the
energy of this QGP string to be about 2.7 GeV/fm for a temperature about
$2T_c$. Lattice simulation of pure gauge theories should be able to 
investigate properties  of these strings. With the interpretation that 
quark contributions lead to explicit breaking of this $Z_N$ 
symmetry \cite{qurk2,psrsk,psrsk2}, 
such {\it QGP} strings may play important role in the evolution of the 
quark-gluon plasma phase and in the dynamics of quark-hadron transition.

As these strings exist in the high temperature phase (the $Z(3)$ symmetry 
being restored at low temperatures), formation and evolution of these 
strings will have unconventional features. The dynamics of a first order
quark-hadron transition for a QGP region infested with these QGP 
strings with confining cores may be very different from the conventional 
scenarios. Even for a second order transition, due to large 
inhomogeneities present in the form of this string network, it is 
possible that the transition may proceed from the strings outward.
It will be interesting to investigate if baryon inhomogeneities can be
generated in this way even if the transition is of second order (or a
cross-over). We hope to address these issues in a future work.

\vskip .2in
\centerline {\bf ACKNOWLEDGEMENTS}
\vskip .1in
 
  We are very thankful to Sanatan Digal, Rajarshi Ray, Supratim Sengupta,
Soma Sanyal, Balram Rai, V. Sunil Kumar, and V.K. Tiwari for useful 
discussions and comments. We especially thank Amit Kundu, B.K. Patra, 
and Aporva Patel for early discussions on this subject. 


\newpage

\begin{figure}[h]
\begin{center}
\leavevmode
\epsfysize=10truecm \vbox{\epsfbox{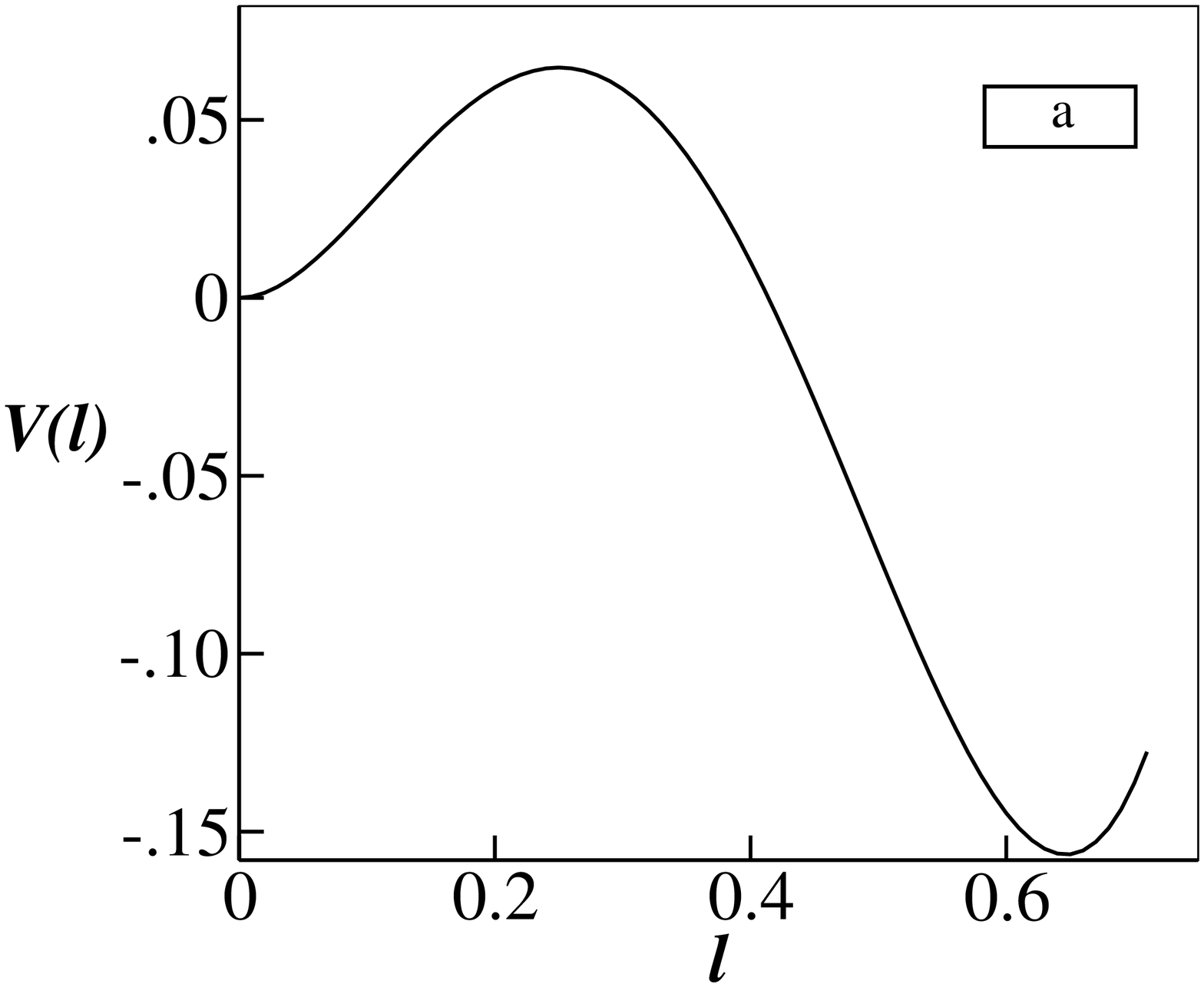}}
\epsfysize=10truecm \vbox{\epsfbox{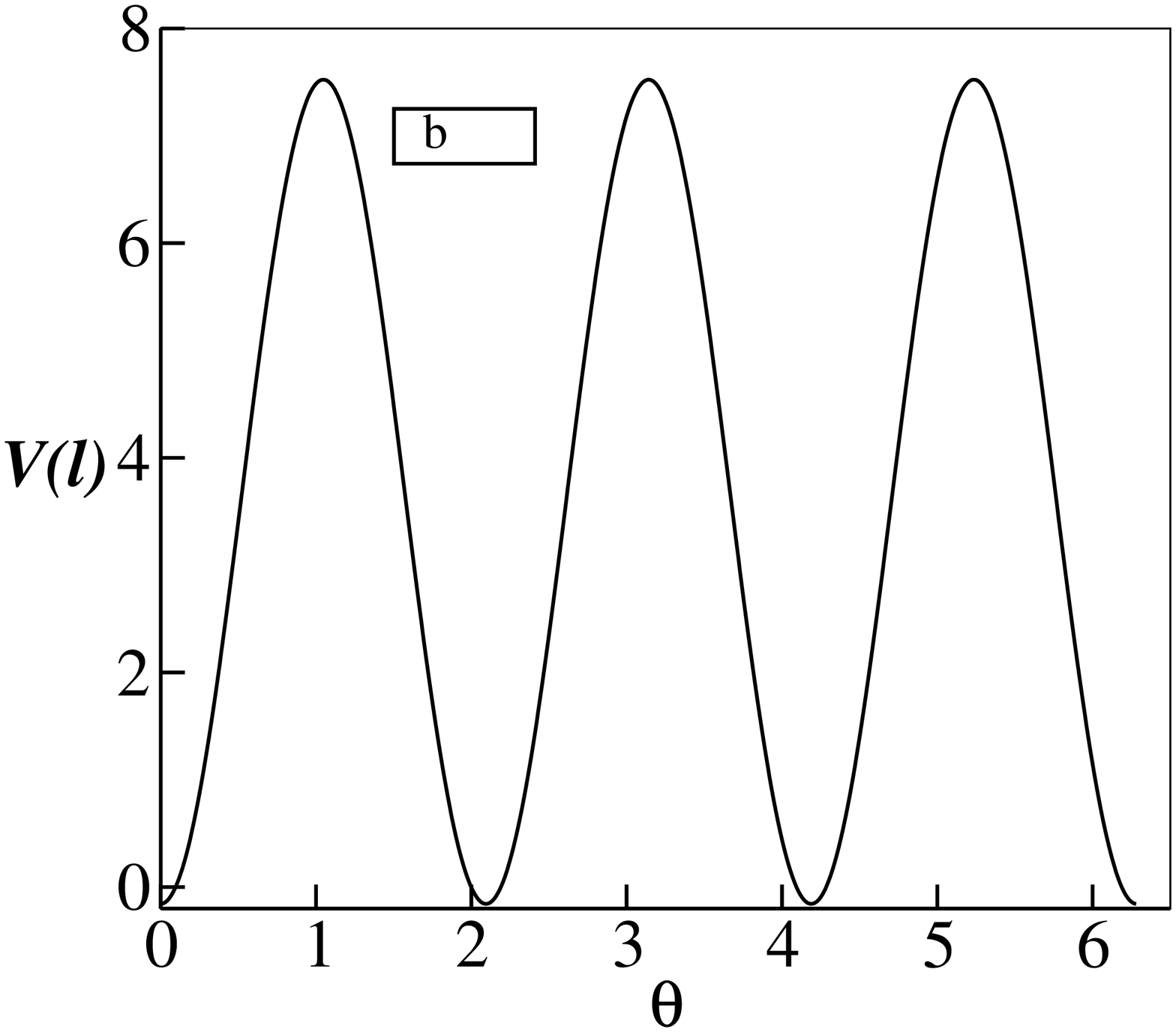}}
\end{center}
\caption{}{(a) shows the plot of  $V(l)$ in $\theta = 0$ direction  
for $T \simeq 185 MeV$. In (a) and (b), plots of $V$ are given in units
of $T_c^4$. The value of critical temperature $T_c = 182 MeV$. 
The plot shows the metastable vacuum at $l = 0$. The $Z(3)$ structure of 
the vacuum  can be seen in (b) in the plot of the potential $V(l)$ as a 
function of $\theta$ for fixed $|l| = l_0$.
Here, $l_0$ corrseponds to the abosute minimum of $V(l)$.} 
\label{Fig.1}
\end{figure}
\newpage

\begin{figure}[h]
\begin{center}
\leavevmode
\epsfysize=20truecm \vbox{\epsfbox{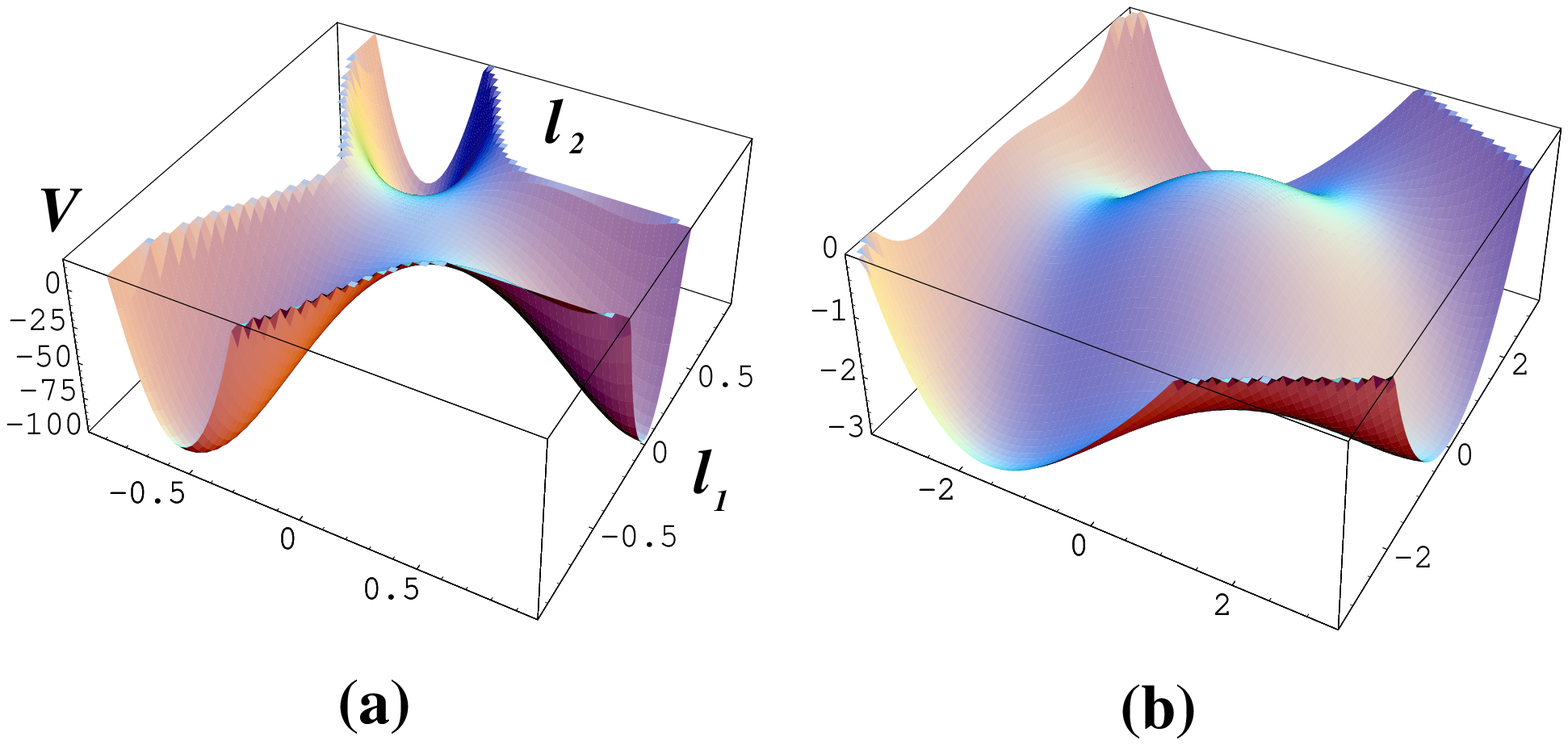}}
\end{center}
\caption{}{(a) shows the surface plot of the potential $V(l)$ (in units
of $T_c^4$) in Eq.(4) in the complex $l$ plane for $T = 400 MeV$. Plot range 
in the $V$ axis is restricted to make the $Z(3)$ barriers and the central bump 
distinctly visible. It is clearly seen that the barrier between different
$Z(3)$ vacua is higher than the central bump. To compare this situation 
with a standard axionic model case (with same $Z(3)$ symmetry), we show 
in (b) the plot of the potential (divided by $10^4$) in Eq.(4) with the
coefficient $b_2$ multiplied by 100. This makes the barrier between
different $Z(3)$ vacua lower than the central bump.}
\label{Fig.2}
\end{figure}

\newpage
\begin{figure}[h]
\begin{center}
\leavevmode
\epsfysize=20truecm \vbox{\epsfbox{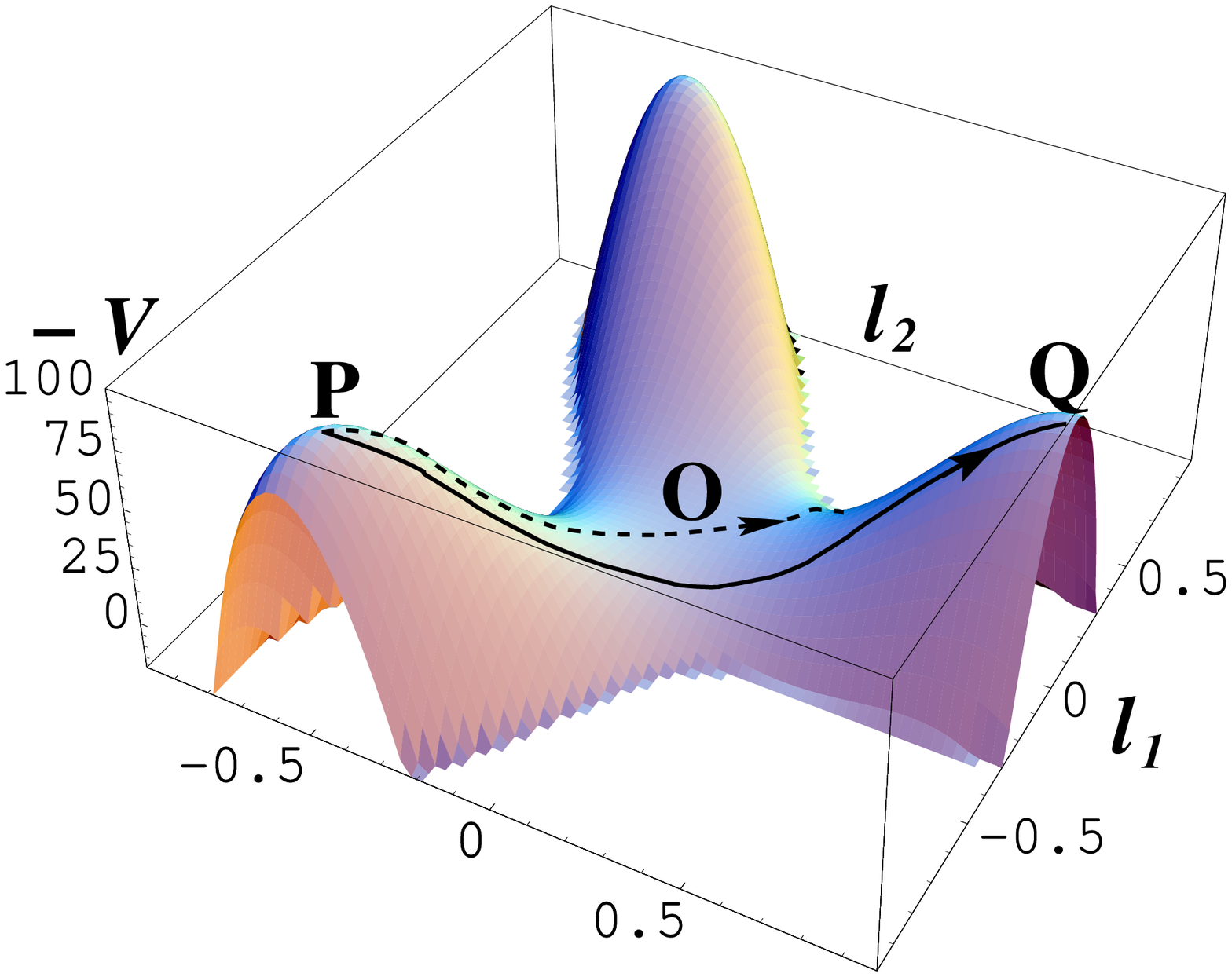}}
\end{center}
\caption{}{Plot of the inverted potential, i.e. $ - V(l_1,l_2)$.
Again, to show the potential shape clearly, plot range is restricted
for negative values. The domain wall solution will correspond to
the particle trajectory starting at the top of one of the hills,
say, at $P$, and ending at the other hill, say at $Q$. 
The patricle starting at $P$, and rolling down to $l = 0$ (point
O in the figure)  cannot turn back to end up at $Q$. It will
rather go to the other side and roll away downwards, as shown by the
dashed curve in the figure. To end up at $Q$, the trajectory must loop back
before reaching $l = 0$, as shown by the solid curve in Fig.3.}
\label{Fig.3}
\end{figure}
\newpage
\begin{figure}[h]
\begin{center}
\leavevmode
\epsfysize=15truecm \vbox{\epsfbox{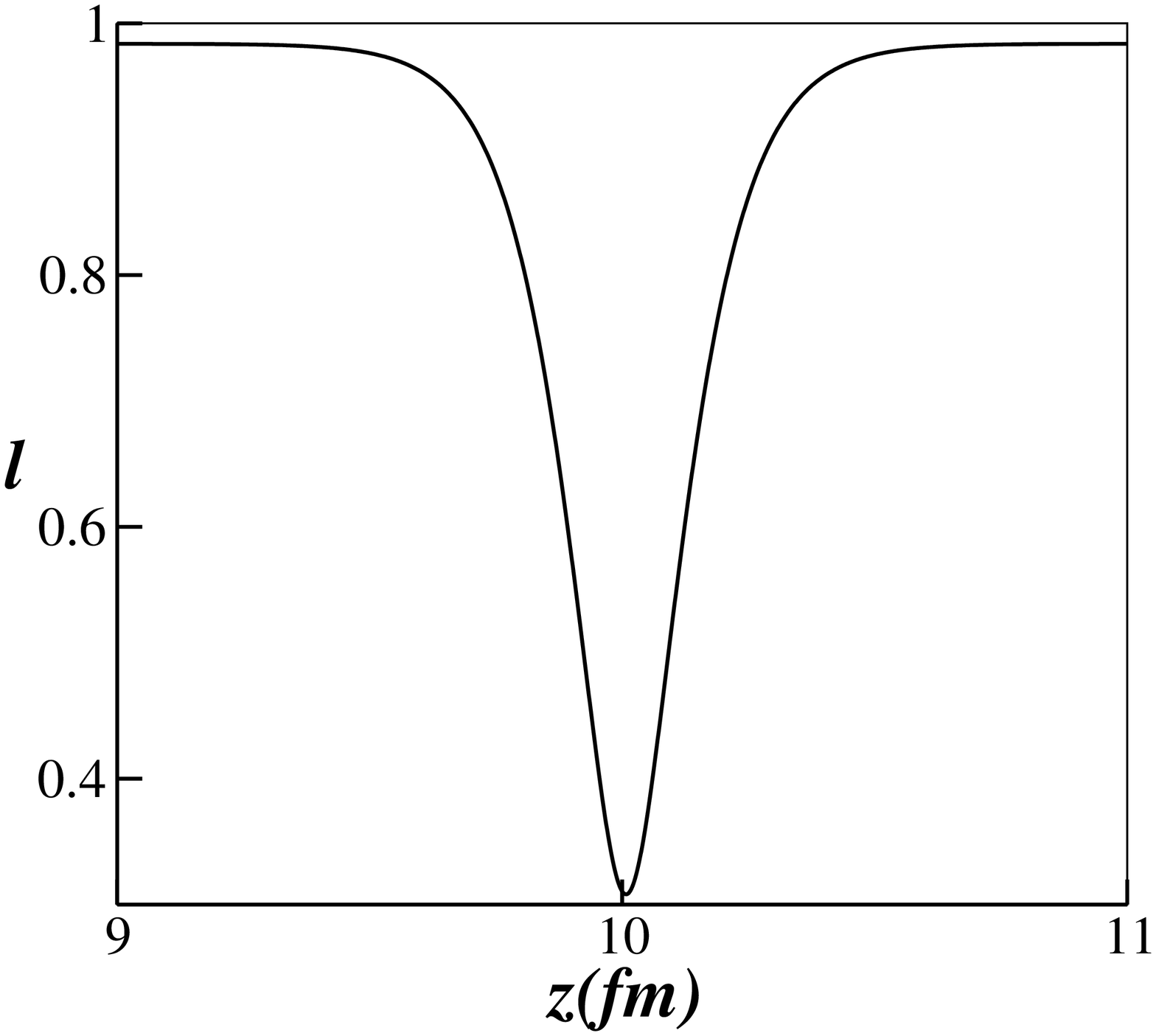}}
\end{center}
\caption{}{The profile for the domain wall solution (centered near
$z = 10$ fm) for $T = 400 MeV$.  Note that $l$ remains non-zero inside
the wall, with the lowest value of $l$ being about 0.3. Wall thickness
(where $l \simeq$ 0.9) is seen to be about 0.5 fm.}
\label{Fig.4}
\end{figure}
\newpage
\begin{figure}[h]
\begin{center}
\leavevmode
\epsfysize=15truecm \vbox{\epsfbox{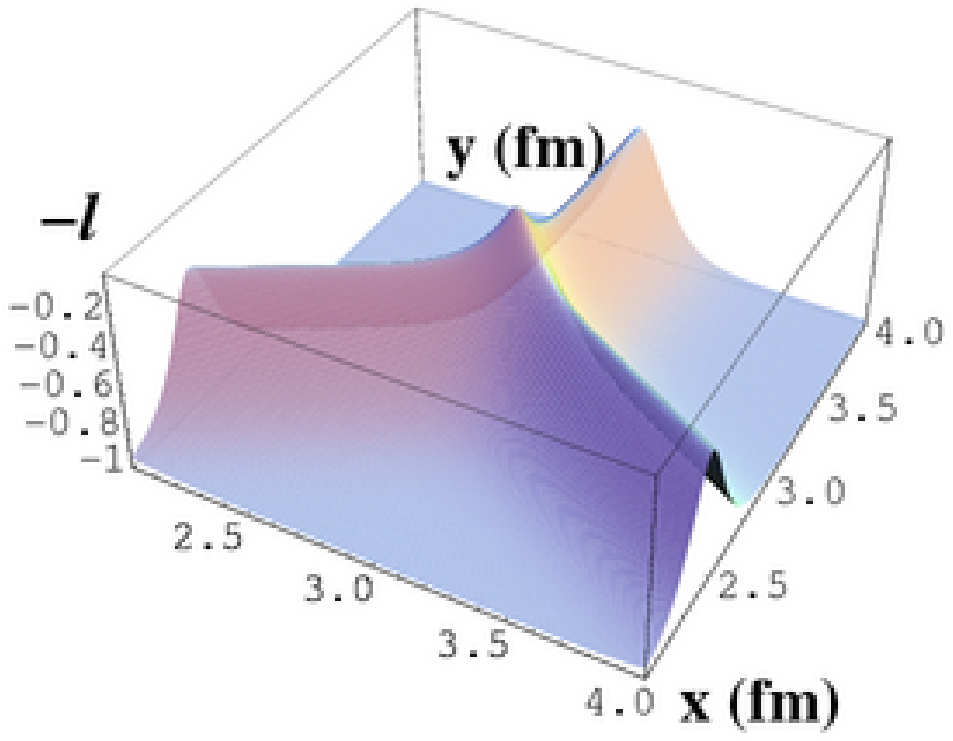}}
\end{center}
\caption{}{Surface plot of $-l$ for a small portion of the two 
dimensional lattice showing clearly the profile of the string,
connected to three interfaces.}
\label{Fig.5}
\end{figure}
\newpage
\begin{figure}[h]
\begin{center}
\leavevmode
\epsfysize=15truecm \vbox{\epsfbox{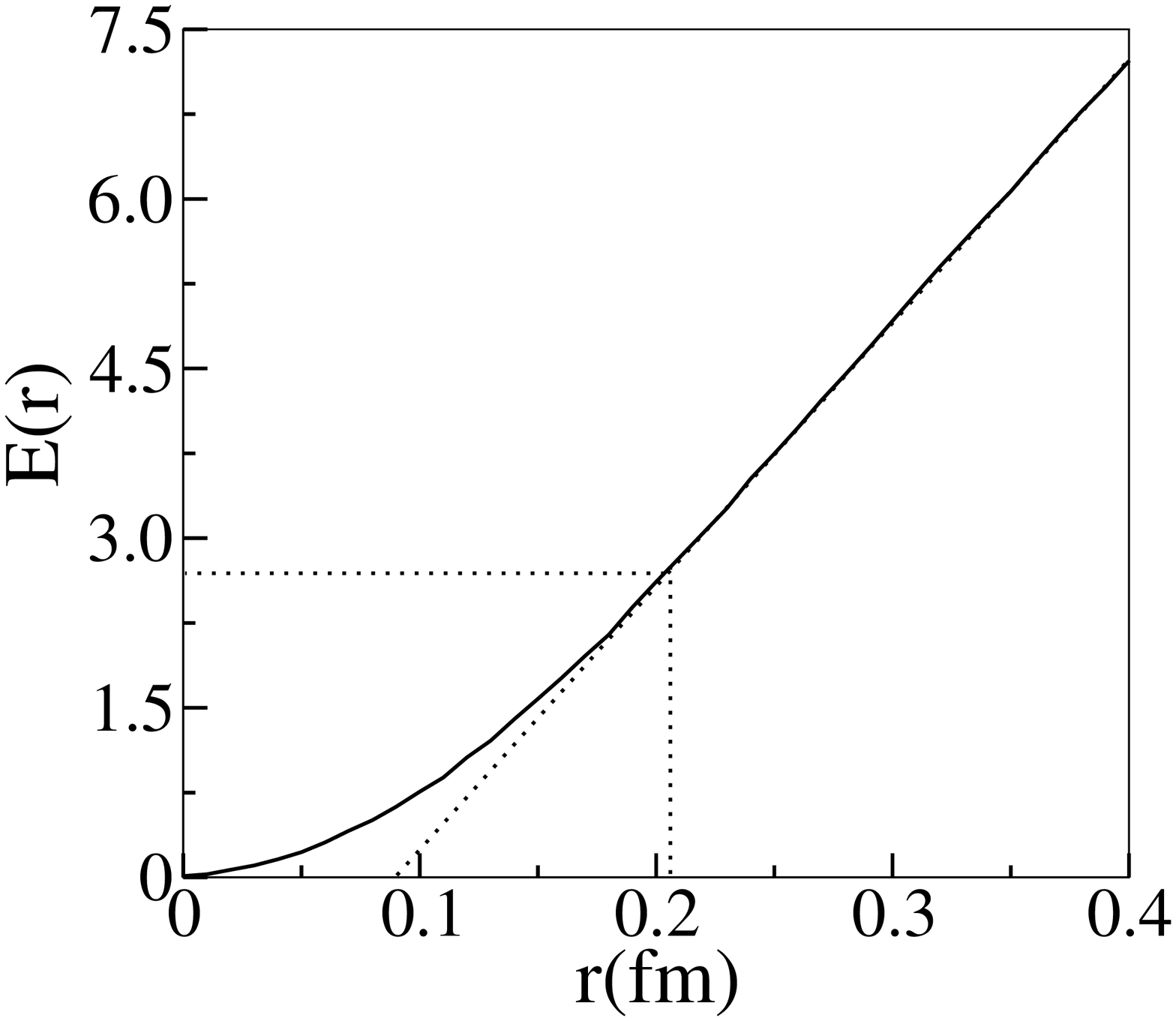}}
\end{center}
\caption{}{Plot of $E(r)$ in GeV/fm vs. $r$. Simulation results are shown 
by the solid curve. Dashed line shows the fitting of the large $r$ part of 
the plot with a straight line.  Its slope is found to be about 23 GeV/fm$^2$ 
giving the value of $\sigma \simeq 7.7$ GeV/fm$^2$.  The 
core energy $E_0(r_0) = E(r_0)$ (Eq.(8)) is identified to be equal to 
$E(r)$ at the value of $r$ where $E(r)$ starts deviating
(for decreasing $r$) from the linear fit shown by the dashed line.
This is shown in the figure, with $E_0$ found to be about 2.7 GeV/fm.}
\label{Fig.6}
\end{figure}
\end{document}